\documentclass[11pt]{article}
\usepackage[utf8]{inputenc}
\usepackage{amsmath}
\numberwithin{equation}{section}
\newenvironment{rcases}
  {\left.\begin{aligned}}
  {\end{aligned}\right\rbrace}
\usepackage{amssymb}
\usepackage{hyperref}
\usepackage{graphicx}
\graphicspath{{./images/}}
\usepackage{float}
\usepackage{multirow}
\usepackage{booktabs}
\usepackage{bigstrut}
\usepackage{gensymb}
\usepackage{geometry}
\usepackage{fancyhdr}
\usepackage{physics}
\usepackage{bm}
\usepackage[dvipsnames]{xcolor}

\usepackage{subfig}

\def\o{\omega}

\def\d{\delta}

\def\p{\partial}
\def\f{\frac}

\usepackage[T1]{fontenc}
\usepackage[utf8]{inputenc}
\usepackage{authblk}

\title{\Large{\bf Hydrodynamics of $(1+1)$ dimensional fluid in the presence of gravitational anomaly from first order thermodynamics}}
\author[1]{Abhinove Nagarajan Seenivasan\thanks{abhinove523@gmail.com, abhinovens@rnd.iitg.ac.in}}
\author[1]{Sayan Chakrabarti\thanks{sayan.chakrabarti@iitg.ac.in}}
\author[1]{Bibhas Ranjan Majhi\thanks{bibhas.majhi@iitg.ac.in}}
\affil[1]{\it Department of Physics, Indian Institute of Technology Guwahati, Guwahati 781039, Assam, India}

\begin{document}
  \maketitle

\begin{abstract}
Considering ($1+1$)-dimensional fluid in presence of gravitational trace anomaly, as an effective description of higher-dimensional fluid, the hydrodynamics is discussed through a first order thermodynamic description. Contrary to the existing approaches which are second order in nature, the fluid velocity is identified through the auxiliary field required to describe the Polyakov action for the effective description of relevant energy-momentum tensor. The thermodynamic and fluid quantities, on a static black hole spacetime, are calculated both near the horizon as well as at the asymptotic infinity. The Unruh vacuum appears to be suitable one for the present analysis, in contrast to Israel-Hartle-Hawking vacuum which is consistent with second order description. As in anomaly cancellation approach to find the Hawking flux the Unruh vacuum is consistent with the required conditions, in reverse way we interpret this fluid description as an alternative approach to find these required conditions to calculate the same in anomaly cancellation approach.
\end{abstract}
\clearpage
\fancyhead[L]{}
\fancyhead[R]{}

\parindent 0 ex

\tableofcontents

\fancyhead[L]{}
\fancyhead[R]{}

\newpage 

\section{Introduction}
Anomalies in field theory at the quantum scale is quite ubiquitous. Usually the classical symmetries may break down in the quantum regime and leads to anomalies. Breaking of diffeomorphism symmetry yields non-conservation of energy-momentum tensor (EMT) and similarly breaking of conformal symmetry generates trace anomaly. In gravitational theories one encounters such scenarios and therefore either trace (for non-chiral case) or both (for chiral case) anomalies appear \cite{alvarez1984gravitational,bardeen1984consistent,BK}.  These anomalies are well studied through effective action and the anomalous stress-tensor can be obtained from its variation with respect to background spacetime metric. The gravitational anomalies for non-chiral theory in $(1+1)$ dimensions is manifested through Polyakov action \cite{polyakov1981quantum} while other one corresponds to Leutwyler action \cite{Leutwyler:1984nd}.  

It has been observed that the fluid constitutive relations get corrected in presence of gravitational anomalies \cite{Landsteiner:2011cp,Dubovsky:2011sk,Valle:2012em,Jensen_2013,Banerjee:2013qha,Banerjee_2014,Majhi_2014,Majhi:2014hsa,Banerjee:2014cya,Banerjee:2014ita}. In $(1+1)$ spacetime, the contribution to the corrections correspond to zeroth order and higher order derivatives of the fluid variables (particularly, see \cite{Jensen_2013,Banerjee_2014,Majhi_2014}). In these analysis the four-velocity components of the fluid is expressed in terms of metric coefficients in the comoving frame. Such a choice yields the anomalous stress-tensor in terms of fluid velocity in which this appears as higher order derivative (more precisely second order derivative terms related to fluid variables) with respect to spacetime. In this direction various aspects, like thermodynamics of such fluid and extension to higher dimensions remain central attention for the last few years \cite{Jensen_2013,Majhi:2014hsa}. It has been observed that inclusion of such corrections in fluid constitutive relations have quiet resemblance with the modification of those through derivative expansion approach. In this description it is worth mentioning that the vacuum states of quantum fields (like Boulware, Unruh, Israel-Hartle-Hawking vacuua) in curved background play pivotal role. Particularly one finds that the Israel-Hartle-Hawking vacuum is the suitable one \cite{Majhi_2014,Majhi:2014hsa}.

In the context of fluid thermodynamics, Eckart's first order formalism \cite{PhysRev.58.919} provides a connection between the thermodynamic variables with the fluid parameters. The original analysis does not consider the diffusion of particles. However a general version has been provided by incorporating the diffusion current \cite{Andersson:2006nr}. This way is a very fruitful one to analyze the thermodynamic aspects of a fluid. Although there are concerns regarding acausal propagation and instabilities in the Eckart model \cite{PhysRevD.31.725,israel1979transient}, we believe that this approach is still worth studying for the following reasons. Firstly, the study of anomalies in ($1+1$)-dimensional hydrodynamics from a thermodynamic perspective is not common, and Eckart's formalism is a good starting point for such an analysis. Importantly, in analysing where the instabilities arise in Eckart's formalism \cite{PhysRevD.31.725}, one finds a crucial contribution that such instabilities comes from components of the dissipative quantities transverse to the (spatial) direction of propagation of the fluid. Since we work in (1+1) dimensional spacetime, these contributions will not arise as there are no spatial transverse directions to the direction of propagation. Hence, we suspect that instabilities and acausal propagation will not be a major cause of concern in (1+1) dimensional spacetime while dealing with Eckart's formalism. In summary, although Eckart's formalism is not a viable one, we take this as a purely mathematical model for its simplicity and want to test whether thermodynamic description of fluid can provide interesting information about anomalous fluid. 
A few recent works \cite{Kovtun_2019, PhysRevD.100.104020} also suggest that first order thermodynamic models are causal and stable if equilibrium states are defined appropriately. Further, although there exist more sophisticated models like the Israel-Stewart formalism \cite{israel1979transient} at second order in the perturbations, computing and comparing the dissipative perturbations in such models are more intensive, and require solving additional dynamical equations for the perturbations. In the same spirit, such an approach has been widely applied to the recent study of scalar fluids in various gravitational theories, like scalar-tensor theory of gravity \cite{Faraoni:2021lfc,Faraoni:2021jri,Giardino:2022sdv,Faraoni:2022jyd,Faraoni:2022doe,Faraoni:2022fxo,Faraoni:2022gry,Faraoni:2023ann,Faraoni:2023hwu,https://doi.org/10.48550/arxiv.2302.00373}, Horndeski gravity \cite{Giusti:2021sku,Miranda:2022wkz,Faraoni:2023hwu}. In these analyses, the fluid four-velocity have been described through the covariant derivative of the scalar field of the theory. It has been observed that the equilibrium situation (namely the general theory of relativity limit) can be described through vanishing of either fluid temperature or its chemical potential. Till now, as far as our knowledge is concerned, anomalous fluid has not been discussed in this setup. Since Eckart's analysis is based on the covariant conservation of EMT, here we intend to investigate such case in presence of only gravitational trace anomaly in $(1+1)$ spacetime dimensions. Our focus will be only the anomalous part of the constitutive relation or the energy-momentum tensor of the fluid.

As mentioned earlier, the anomalous stress-tensor can equivalently be obtained through effective action. For our present study it is given by Polyakov action \cite{polyakov1981quantum}. Originally such an action appears in non-local form, which can be cast into local form by introducing an auxiliary scalar field. Variation of this form with respect to metric tensor provides the anomalous stress-tensor which comes in terms of first order as well as higher order derivatives of the scalar field and is covariantly conserved. In the local form formalism, the scalar field satisfies Klein-Gordon type equation which is sourced by Ricci scalar of the background metric. In this case, the non-vanishing trace of EMT is proportional to the Ricci scalar. This local version of EMT provides a fluid description in presence of the particular gravitational anomaly. In the present analysis, we wish to define the velocity of fluid in terms of the derivative of the auxiliary scalar field. Using such identification and the Eckart's formalism, all the fluid variables and thermodynamic parameters are obtained. In order to have better understanding of these quantities, we find the explicit forms in terms of the metric coefficients for a static background which is associated with a Killing horizon. We observe that among three relevant vacua, Unruh vacuum is the suitable one in this description. This is in sharp contrast with the earlier higher order description. The fluid and thermodynamic parameters are finite and non-vanishing near the Killing horizon, while they vanish at the asymptotic spatial infinity. This is consistent with the fact that the anomaly vanishes at infinity while it does not vanish near the horizon.

Through this analysis we observe an interesting connection between the calculation of the Hawking flux through the imposition of Unruh vacuum on the components of the EMT (known as the anomaly approach, please see, e.g. \cite{Christensen:1977jc,Balbinot_1999} through trace anomaly and \cite{Robinson:2005pd,Iso:2006wa,Banerjee:2007qs,Banerjee:2007uc,Banerjee_2009,Banerjee_2009_02} for chiral case) and the present application of Eckart's formalism to identify the fluid variables. The relevance of Unruh vacuum in both indicates a close connection between them. We find that imposition of certain conditions on the fluid parameters equally capable of mimicking the imposition of Unruh vacuum condition on the EMT. Since anomaly vanishes at infinity, one demands that the anomalous fluid parameters (particularly the heat conductivity and the viscosity) must vanish. On the other hand, they have to be finite at the horizon. Interestingly, these demands are sufficient to yield the required conditions which are consistent with the Unruh vacuum and so with the ingredients of anomaly cancellation technique. 
Therefore, we feel that the fluid description of the anomalous EMT within the Eckart's formalism with the identification of the velocity through the covariant derivative of auxiliary scalar field, can have connection with the anomaly cancellation approach.

The organization of the paper is as follows.  In Section \ref{review} we briefly review the action and the EMT for the anomalous scalar fluid, and discuss the first order thermodynamics for a relativistic fluid. We then compute the fluid and thermodynamic parameters for this fluid in Section \ref{parameters}. The fluid and thermodynamic parameters are considered for the special case of a spherically symmetric, static background in Section \ref{staticbg}. For such a background, we consider the EMT of the fluid, in three different vacuum states in Section \ref{EMvacua}. In Section \ref{UnruhFluid}, we discuss in detail the properties of this fluid in the Unruh vacuum state and explain a novel connection with Hawking's effect in Section \ref{Hawking}. We conclude our work with Section \ref{conclusion}. 

\section{Anomalous fluid and first order thermodynamics: a brief review} \label{review}
This section reviews the fluid description in presence gravitational anomaly in $(1+1)$ spacetime dimensions. Also we present a brief summary of first order thermodynamics, specifically Eckart's formulation to relate the thermodynamics of fluid with the fluid parameters. These will build the basics of our main discussion.

\subsection{Gravitational anomaly and anomalous fluid}
Anomalies are an important aspect of quantum field theory which arise when symmetries in a classical theory do not extend to quantum theory. In this work, we consider gravitational anomalies and restrict our discussion on $(1+1)$ dimensions. These anomalies can manifest in two ways, which can be described through the renormalised energy momentum tensor $T_{ab}$ of the field. These are given by \cite{alvarez1984gravitational,bardeen1984consistent,BK,Jensen_2013, Banerjee_2009, Banerjee_2009_02},
trace anomaly (violation of conformal invariance i.e. $T_a^a = c_wR$, but $\nabla_aT^{ab} = 0$) alone for non-chiral theories, or both trace and
 diffeomorphism anomalies (i.e. violation of conformal invariance as well general covariance, hence $T^a_a = (c_w/2)R$ and $\nabla_bT^{ab} = c_g\bar{\epsilon}^{ab}\nabla_bR$) for chiral theories. Here $c_g, c_w$ are constants, while $R$ is the two dimensional Ricci scalar.  For non-chiral theory the constant is given by $c_w = 1/(24\pi)$ (see section III of \cite{Christensen:1977jc}), while for chiral one the constants are as follows: $c_w=1/(48\pi)$ and $c_g = 1/(96\pi)$ \cite{BK}. In a non-chiral theory, one can generally retain diffeomorphism symmetry at the cost of a trace anomaly. In the present analysis, we consider only non-chiral one. The reason is -- we will use a formalism (will be briefed in the next-subsection) in which the covariant conservation of EMT is a necessary ingredient.  
 
 Usually the anomaly shows up in the effective action. In our case, the anomaly induced effective action is the well known non-local form of Polyakov action \cite{polyakov1981quantum}:
\begin{equation}
S^{(g)}_{\text{eff}} = \f{c_w}{4}\int d^2x~d^2y~\sqrt{-g}R(x)\f{1}{\Box}(x,y)\sqrt{-g}R(y)~.
\end{equation}
In the above $1/\Box$ is the inverse of the Laplacian $g^{ab}\nabla_a\nabla_b = (1/\sqrt{-g})\p_a\left(\sqrt{-g}g^{ab}\p_b\right)$. This non-local form can be converted to a local one by introducing an auxiliary field $\Phi$. The action is then given by
\begin{equation}
S_P^{(g)} = \f{c_w}{4}\int d^2x~\sqrt{-g}\left(-\Phi\Box\Phi + 2R\Phi\right)~,
\label{PRDV41}
\end{equation}
where $\Phi$ satisfies the following equation
\begin{equation}
\label{phieom}
\Box\Phi = R~.
\end{equation}
Before giving the expression for EM tensor, let us make a side comment on the structure of the effective action (\ref{PRDV41}). It may be noted that the first term in the action can be written as $-\Phi\Box\Phi = \nabla_a\Phi \nabla^a\Phi - \nabla_a(\Phi\nabla^a\Phi)$. Therefore, except the unimportant total derivative term, this has a similarity with the action of a massless phantom scalar field.

The anomalous energy-momentum tensor is then evaluated as
\begin{equation}
\label{EMTPhi}
T_{ab} = -\f{2}{\sqrt{-g}}\f{\d S^{(g)}_{P}}{\d g^{ab}} = \f{c_w}{2}\left\{\nabla_a\Phi\nabla_b\Phi - 2 \nabla_a\nabla_b\Phi + g_{ab}\left(2 R - \f{1}{2}\nabla_e\Phi\nabla^e\Phi\right)\right\}~.
\end{equation}
It can be verified that the above form of the EMT satisfies
\begin{equation}
 T^a_a = c_w R~,
 \label{Trace}
\end{equation}
with $\nabla_a T^{ab} = 0$ and $c_w = 1/(24\pi)$. Note that in (\ref{EMTPhi}) the coefficient $c_w$ appears as an over all multiplicative factor and therefore the obtained fluid parameters, as we will see later, also contain this as an over all multiplicative factor. However, we will continue to use this without explicitly substituting its numerical value, except wherever necessary.

In the literature, extensive discussions exist to find the corrections to the energy-momentum tensor of a fluid due to these anomalies (e.g. see \cite{Valle:2012em,Jensen_2013,Banerjee:2013qha,Banerjee_2014,Majhi_2014,Banerjee:2014cya,Banerjee:2014ita}). It is to be noted that, the existing analysis is mainly based on a particular way of identification of fluid velocity in terms of the gravitational parameters. It has been observed that the constitutive relation incorporates higher order derivative corrections of fluid four-velocity. Consequently, the fluid parameters, like pressure, energy density etc., incorporate higher derivative corrections related to zeroth order fluid parameters. The analysis may be extended to find the existence of both types of anomalies as well as to $(3+1)$ spacetime dimensions. Particularly, in the formalism for non-chiral theory in $(1+1)$ dimensions, it has been observed that the anomalous constitutive relation (\ref{EMTPhi}), when written in terms of the fluid four velocity, does not incorporate viscous stress-tensor \cite{Jensen_2013}, but at the same time it is also not same as that of the ideal fluid. The reason being the energy-momentum tensor incorporates higher derivative terms due to presence of gravitational anomaly (see also \cite{Banerjee_2014}). This method is usually known as derivative expansion approach. 

Contrary to the existing approach, in this paper, we aim to understand the thermodynamics as well as fluid properties corresponding to the anomalous fluid, represented by (\ref{EMTPhi}), through a first order formalism, such as the one proposed by Eckart \cite{PhysRev.58.919}. Therefore in the next sub-section a brief review on (extended) Eckart's formulation \cite{Andersson:2006nr} will be presented. We note here that these results may be considered to be in a general frame, as opposed to the usual notions of Eckart or Landau frame, based on the choice of four-velocity and number current. This is a well established approach existing in the literature as well \cite{Andersson:2006nr}.

\subsection{Eckart's First Order Thermodynamics}
A framework to understand the (first order) thermodynamics of fluids is given by Eckart's formalism \cite{PhysRev.58.919}. This was originally formulated in $(3+1)$ dimensions. In this case diffusion of particles were not considered. Later on a much generalised version have been proposed where the diffusion current has been incorporated (see e.g. \cite{Andersson:2006nr}). We call this as extended-Eckart's formalism.

 Consider a fluid with four velocity $u_a$ in a background manifold $\left(\mathcal{M}, g_{ab}\right)$. The four-velocity is chosen with the timelike normalisation $u_au^a = -1$ (choosing unit such that $c=1$). 
With our choice of the fluid four-velocity, we can foliate our spacetime into stacks of spacelike hypersurfaces whose metric is given by, 
\begin{equation}
\label{indmet}
h_{ab} = g_{ab} + u_au_b~. 
\end{equation}
The acceleration $\dot{u}_a = u^b\nabla_bu^a$ is orthogonal to the velocity i.e. $\dot{u}_au^a = 0$ and hence spacelike. 

A general imperfect fluid is then described by the following energy-momentum tensor
\begin{equation}
\label{emfull}
T_{ab} = \underbrace{\rho u_au_b + p_{\text{iso}}h_{ab}}_{T_{ab}^{\text{ideal}}} + \underbrace{q_au_b + q_bu_a}_{T_{ab}^{\text{heat}}} + \underbrace{p_{\text{vis}}h_{ab} + \pi_{ab}}_{T_{ab}^{\text{vis}}}~,
\end{equation}
which is covariantly conserved; i.e. $\nabla_a T^{ab} = 0$. Invoking the scalar field/fluid analogy \cite{Piattella_2014, https://doi.org/10.48550/arxiv.2302.00373}, we generally expect dissipative effects in the fluid to arise from terms containing higher derivatives of the scalar field in the energy momentum tensor. The salient features of (extended) Eckart's formalism for an imperfect fluid are summarised below for the reader: 
\begin{align}
q^a &= - \kappa h^{ab}\left(\nabla_bT + \dot{u}_bT\right)~; 
\label{qeck}
\\ 
\pi^{ab} &= - 2 \eta \sigma^{ab}~;
\label{pieck} 
\\ 
p_{\text{vis}} &= - \zeta \theta~;
\label{pviseck} 
\\
\nu^a &= -\sigma T^2 h^{ab}\nabla_b\Big(\frac{\mu}{T}\Big)~.
\label{nu}
\end{align}
Here $q^a$ is the heat flux density, $\kappa$ is heat conductivity, $\sigma$ is the diffusion coefficient, $T$ is the temperature of the fluid, and $\pi^{ab}$ is the viscous stress tensor. $\eta, \zeta$ are called the first and second coefficients of viscosity, and we normally take the bulk/second coefficient of viscosity to be zero. This is because we want to use Eckart's formalism and such is developed with the consideration of $p_{\textrm{vis}}=0$ \cite{PhysRev.58.919}. Consequently we have $\zeta = 0$ and therefore in the subsequently analysis we will set this to zero. $\theta$ is the expansion scalar and $\sigma^{ab}$ is the shear tensor. Also remember that $\kappa, \sigma \geq 0$ and $\nu^a$ is the diffusion flux. The above relations show that the left hand parameters are related with the first order derivative of the fluid and thermodynamic quantities. For this reason the present approach is called first order formalism. Note that these thermodynamic quantities can be obtained by taking suitable projections from the energy momentum tensor: 
\begin{eqnarray}
q^a = - T_{bc}u^bh^{ac}~; \rho = T_{ab}u^au^b~; 
\Pi_{ab} = p_{\text{iso}}h_{ab} + \pi_{ab} = T_{cd}h_a^ch_b^d~; p_{\text{iso}} = T_{ab}h^{ab}~.
\label{B1}
\end{eqnarray}
We give an extensive discussion on the above structure in Appendix \ref{BD1} (following \cite{Andersson:2006nr}) for making the symbols and underlying inputs clear to the reader. This will make the above discussion self-sufficient.
We now intend to use them for (\ref{EMTPhi}) to find these quantities. However, (\ref{EMTPhi}) refers to the fluid in ($1+1$)-dimensions. Therefore a clear justification must be given for using the above viscous description in the present study, which we mention below.

In this work, we study the thermodynamics of a ($1+1$) dimensional relativistic fluid with a trace anomaly. While such fluids are important on their own merit, there also exist in the literature, problems where ($3+1$) dimensional fluids can be studied via an ``effective'' ($1+1$) dimensional fluid description. A typical example is the case of nuclear collision reactions \cite{bjorken1983highly, fries2008stress, li2010stability}, where the quark gluon plasma is a ($3+1$)-dimensional fluid. However, the nature of these collisions is such that the motion of the fluid particles is largely along a single spatial direction with negligible transverse propagation \cite{bjorken1983highly}. Thus, their behaviour can be studied by simply considering one spatial dimension and hence, one can conclude that this ($1+1$)-dimensional fluid will exhibit properties of the underlying ($3+1$)-dimensional fluid, such as the viscosity or shear. Importantly, in this approach, we can continue to use the EM tensors and the framework developed for ($3+1$)-dimensional fluid to describe our effective ($1+1$)-dimensional fluid. The structure of the EM tensor or the interpretation of fluid quantities like the viscosity are those of the underlying ($3+1$)-dimensional fluid. 
In our analysis as well, we follow the same spirit to study a ($1+1$)-dimensional fluid. We assume that the two dimensional fluid in question is an effective description of the yet to be known higher dimensional viscous fluid. We can then use an EM tensor whose structure is identical to that of an imperfect fluid in ($3+1$)-dimensions, and work with fluid and thermodynamic quantities in the framework of this fluid. In that sense (\ref{EMTPhi}) can be considered to be in the form (\ref{emfull}), but now in two spacetime dimensions and therefore can be worked with expressions given in (\ref{B1}). Therefore the obtained description, which we will find, refers to an effective description of an original higher-dimensional one.

\section{Fluid and Thermodynamic Parameters} \label{parameters}
We will use Eckart's identifications to compute the relevant fluid and thermodynamic quantities for the anomalous fluid we consider. First, we define the fluid velocity through the gradient of the auxiliary scalar field $\Phi$, with a suitable time-like normalization i.e. 
\begin{equation} 
\label{4vel}
u_a = \f{\nabla_a \Phi}{\sqrt{-\nabla_i\Phi\nabla^i\Phi}}~, \,\,\,\  u_au^a = - 1~.
\end{equation}
Such a definition of fluid velocity is different from what was used earlier in \cite{Valle:2012em,Jensen_2013,Banerjee:2013qha,Banerjee_2014,Majhi_2014,Majhi:2014hsa,Banerjee:2014cya,Banerjee:2014ita} to describe the properties of anomalous fluid.
Using (\ref{indmet}), (\ref{B1}) and (\ref{4vel}) in (\ref{EMTPhi}) we obtain the parameters as follows: 
\begin{align}
    \rho &= \f{c_w}{2}\left[ - \f{1}{2}\nabla_e\Phi\nabla^e\Phi - 2 R + \f{2\nabla^a\Phi\nabla^b\Phi\nabla_a\nabla_b\Phi}{\nabla_e\Phi\nabla^e\Phi}\right] 
\label{rhoed}~, \\ 
    q_a &= c_w \f{\nabla^b\Phi}{(-\nabla_e\Phi\nabla^e\Phi)^{3/2}}\left[\nabla^c\Phi \nabla_b\nabla_c\Phi\nabla_a\Phi - \nabla_c\Phi\nabla^c\Phi\nabla_b\nabla_a\Phi\right] \label{heatq}, \\ 
    \Pi_{ab} &= \f{c_w}{2}\left[h_{ab}\left(2R - \f{\nabla_e\Phi\nabla^e\Phi}{2}\right) - 2 h_a^ph_b^q\nabla_p\nabla_q\Phi \label{Isostress}\right] , \\
    p_{\text{iso}} &= \f{c_w}{2}\left[\f{2\nabla^a\Phi\nabla^b\Phi\nabla_a\nabla_b\Phi}{\nabla_e\Phi\nabla^e\Phi} - \f{\nabla_e\Phi\nabla^e\Phi}{2} \right], \label{S2}\\ 
    \theta &= \f{1}{\sqrt{-\nabla_i\Phi\nabla^i\Phi}}\left[R - \f{\nabla^a\Phi\nabla^b\Phi\nabla_a\nabla_b\Phi}{\nabla_e\Phi\nabla^e\Phi} \right],\label{BD4} \\ 
    \eta &=  \f{c_w}{2}\sqrt{-\nabla_e\Phi\nabla^e\Phi}~,
    \label{B2}
\end{align}
where in calculating $p_{iso}$ we have used the equation of motion (\ref{phieom}) for $\Phi$. 

As discussed in Appendix \ref{BD1}, it may be pointed out that the whole theory depends on the structure of $\Pi_{ab}$ as an input. We can observe that in this case $\Pi_{ab}$ depends on the first order derivative of velocity (see Eq. (\ref{piab})), and hence so for $\theta$ $(\theta = \nabla_au^a)$ also. We now investigate whether the above expressions for these quantities are consistent with this when $u^a$ is identified as (\ref{4vel}). For the given (\ref{4vel}) it is easy to obtain
\begin{equation}
\nabla_b u_a = \frac{1}{\sqrt{-\nabla_i\Phi\nabla^i\Phi}} \Big[\nabla_a\nabla_b\Phi - \frac{\nabla_a\Phi \nabla^c\Phi \nabla_b\nabla_c\Phi}{\nabla_e\Phi\nabla^e\Phi}\Big]~.
\label{BD3}
\end{equation}
One can then check that $h^{ab}\nabla_bu_a = g^{ab}\nabla_bu_a$ yields (\ref{BD4}). Therefore $\theta$, given by (\ref{BD4}), depends only on the gradient of the identified velocity (\ref{4vel}). Next using (\ref{BD3}) and (\ref{B2}) one finds
\begin{equation}
-\eta h_{ap}h_{bq} (\nabla^qu^p + \nabla^p u^q) = \frac{c_w}{2} \Big[-2 h_{ap}h_{bq}\nabla^q\nabla^p\Phi + h_{ap}h_{bq} \frac{\Big(\nabla^p\Phi \nabla^c\Phi \nabla^q\nabla_c\Phi + (p\leftrightarrow q)\Big)}{\nabla_e\Phi\nabla^e\Phi}\Big]~.
\label{S1}
\end{equation}
Use of (\ref{4vel}) and $u^ah_{ab}=0$ yield the vanishing of the last term. Then substitution of the above in (\ref{piab}) and use of (\ref{S2}) gives us the expression for $\Pi_{ab}$, shown above. Therefore the above $\Pi_{ab}$ is consistent with the expression (\ref{piab}) and hence our identified $\Pi_{ab}$ is function of velocity gradient. Hence $\Pi_{ab}$ can be expressed as
\begin{eqnarray}
\Pi_{ab} &=& \pi_{ab}+p_{\text{iso}}h_{ab}
\nonumber
\\
&=& - \eta \left\{h_{a p}h_{b q} \left(\nabla^{q}u^{p} + \nabla^{p}u^{q}\right) - {2}h_{ab}\nabla_{p}u^{p}\right\} +p_{\text{iso}}h_{ab}~.
\label{S5}
\end{eqnarray}
Consequently we can check that $p_{\text{iso}}=\Pi^a_a$ yields the above given value in terms of $\Phi$. The discussion provided above, verifies the consistency of our results for fluid parameters in terms of $\Phi$.

Particularly, note that given the form of the velocity in (\ref{4vel}), one can also obtain the acceleration in terms of the scalar field. This is given by,
\begin{equation}\label{4accln}
 a_a = \dot{u}_a = \left(-\nabla_e\phi~\nabla^e\phi\right)^{-2}\nabla^b\phi\left[\left(-\nabla_e\phi~\nabla^e\phi\right)\nabla_b\nabla_a\phi + \nabla^c\phi\nabla_b\nabla_c\phi\nabla_a\phi\right]~,
\end{equation}
and is a well known relation provided in the literature \cite{Faraoni:2021jri, Faraoni:2021lfc, Faraoni:2022gry}. We also observe from (\ref{4vel}) and (\ref{4accln}) that the heat flux $q_a$ in (\ref{heatq}) is proportional to the acceleration, and can be written as, 
\begin{equation}
q_a = c_w~\sqrt{-\nabla_e\Phi\nabla^e\Phi}~\dot{u}_a~.
\end{equation}
However, comparing this with the standard form in (\ref{qeck}) we expect the heat flux to be $q^a = - \kappa h^{ab}\left(\nabla_bT + \dot{u}_bT\right)$. Thus, we can identify the following parameters from our expression for $q_a$ in terms of the scalar field:
\begin{equation}
\label{kappaT}
\kappa T = - c_w~\sqrt{-\nabla_e\Phi\nabla^e\Phi}
\end{equation}
with the condition, 
\begin{equation}
\label{spatialder}
\kappa h_{ab}\nabla^bT = 0~.
\end{equation}
Note that this condition is a consequence of the fact that our heat flux turns out to be proportional to the acceleration. This method of identifying our parameters has been followed from the literature \cite{Faraoni:2021jri, Faraoni:2021lfc, Faraoni:2022gry}. In this context, we expect that the thermal conductivity $\kappa$ is determined indirectly by the scalar field and/or by the velocity (derivatives of the scalar field). Hence, one cannot absorb $\kappa$ into the scalar field or vice versa. Moreover we will keep $\kappa>0$ and hence our identified temperature will come out to be negative one. More about this will be discussed in the later analysis. Also note that in this case the obtained $q_a$ in (\ref{heatq}) is of the form $q_a = -\kappa T {\dot{u}}_a = -\kappa T u^b\nabla_b u_a$. Hence it, as expected, depends on the gradient of $u_a$.

This ends the discussion on the thermodynamic description of our anomalous fluid through Eckart's formalism. Again, we point out that the viscous quantities obtained here are to be interpreted in the context of an underlying (3+1) dimensional imperfect fluid, whose effective description is provided by the (1+1) dimensional fluid under study. In order to understand the properties and behaviour of these quantities, in the next section, we will analyse them under a static background.


\section{Static background in $(1+1)$ dimensions} \label{staticbg}
To understand the properties of the quantities (\ref{rhoed}) -- (\ref{B2}) and their feasibility and/or viability, we prepare the simplest background in which they will explicitly be evaluated. Then it will be easier for us  to know about these quantities in detail. 
For ease of computation, we choose a static, background in $(1+1)$ dimensions, given by: 
\begin{equation}
\label{rtmet}
ds^2 = - f(r)~dt^2 + \f{1}{f(r)}dr^2~.
\end{equation}
The metric is static and has a time-like Killing vector corresponding to $\p/\p t$ with the Killing horizon $r_H$ given by the solution of the equation $g_{00}\vert_{r_H} = f(r)\vert_{r_H} = 0$. We can also work in null coordinates $(u,v)$ which are given by
\begin{equation}
u = t - r_*~; v = t + r_*~,
\end{equation}
where the tortoise coordinate $r_*$ is related to the usual radial coordinate as $dr_* = dr/f(r)$. Note that in these coordinates, the metric only has non-zero off-diagonal components i.e. 
\begin{equation}
\label{uvmet}
ds^2 = -\f{f}{2} \left(dudv + dvdu\right)~.
\end{equation}

For the dynamical equation $\Box \Phi = R$ in this background, we have the solution \cite{Banerjee_2014}
\begin{equation}
\Phi = \Phi_0(r) - 4 pt + q~;\p_r\Phi_0 = \f{z-f'}{f}, 
\label{PRDV31}
\end{equation}
where the prime denotes derivatives with respect to $r$, and $p,q,z$ are constants. To determine these constant we need other physical boundary conditions. For a static and asymptotically flat spacetime, there are well defined vacuum states for quantum fields. We will impose those vacuum conditions to fix them.

\subsection{Thermodynamic and Fluid Parameters}\label{thermofluidpar}
To compute the relevant thermodynamic and fluid quantities, we first identify $N_0 \equiv \nabla_i\Phi\nabla^i\Phi$ and $M_0 \equiv \nabla^a\Phi\nabla^b\Phi\nabla_a\nabla_b\Phi$ (see Appendix \ref{App01}). Recall that $\kappa T$ is given by (\ref{kappaT}) such that $\kappa \equiv \kappa (r)$ satisfies (\ref{spatialder}). Since for static spacetime $T$ is only a function of $r$, we must have:
\begin{align}
\kappa h_{ab} g^{bc}\nabla_cT &= \kappa h_{rr} g^{rr} \p_rT =0~.
\end{align}
Then, the above condition yields two possibilities (see Appendix \ref{App 02}): 
\begin{equation}\label{2choices}
p = 0~, \hspace{7 px} \text{or} \,\,\ \hspace{17 px} \kappa = \sqrt{\f{16p^2 - (z-f^{\prime})^2}{f}}~.
\end{equation}

For the choice $p=0$, one has $N_0 = \f{(z-f')^2}{f}$ (see Eq. (\ref{N0})), which is a positive quantity for $r>r_H$. Then the normalization factor $\sqrt{-N_0}$ in the expression of fluid velocity (see Eq. (\ref{4vel})) is not well defined. This means the velocity vector will face issues. This is also evident from the fact that $u^t$ vanishes for $p=0$ and therefore $u^a$ becomes spacelike vector. Hence the choice $p=0$ should be discarded in this analysis. Then, we will consider only the second solution for the later discussion. Now that we have an explicit form for $\kappa$, we can also compute the temperature using (\ref{kappaT}), which yields $T = -c_w$. This is a constant negative temperature. A detailed discussion regarding the physical significance of such a temperature and when it can exist is given in subsection (\ref{UFHorizon}).

Using this, we can write down all thermodynamic quantities in terms of the metric functions and the constants, which are given below:
\begin{align}
    \rho &= \f{c_w}{4}\left[\frac{64 p^2 f f''+\left(32 p^2 z-2 z^3\right) f'+2 z (f')^3-(f')^4+\left(z^2-16 p^2\right)^2}{f \left(16p^2 - (z-f')^2\right)}\right]~; \label{B3}\\ 
    p_{iso} &= \f{c_w}{4}\left[\frac{(f')^2+16 p^2-z^2}{f}+\frac{4 f'' \left(z-f'\right)^2}{\left(-f'+4 p+z\right) \left(f'+4 p-z\right)}\right]~;\\ 
    \kappa &= \sqrt{\f{16p^2 - (z-f^{\prime})^2}{f}}~; \label{B10}\\ 
    \eta &= \f{c_w}{2}\sqrt{\f{16p^2 - (z-f^{\prime})^2}{f}}~;\label{B11}\\ 
    \theta &= \frac{-32 p^2 f f''+\left(16 p^2 z-z^3\right) f'+\left(3 z^2-16 p^2\right) (f')^2-3 z (f')^3+(f')^4}{2 \sqrt{f} \left(2 z f'-(f')^2+16 p^2-z^2\right)^{3/2}}~.
\label{B4}
\end{align}

We can also compute the norm of the heat flux density $q_aq^a$. This is given by, 
\begin{align}
    q_aq^a &= \frac{\Big[p f' \left(-2 z f'+f'^2-16 p^2+z^2\right)+2 p f f'' \left(z-f'\right)\Big]^2}{576 \pi ^2 f^2 \Big[-2 z f'+f'^2-16 p^2+z^2\Big]^2} \label{qaqa}
\end{align}
Note that the norm is positive, as expected for spacelike $q_a$.

\section{EM tensor in different vacua}\label{EMvacua}
Before proceeding toward the analysis of the quantities defined in (\ref{rhoed}) -- (\ref{B2}), we will briefly discuss the values of the constants $p,q,z$ under different choices of vacuum states. The well defined vacua are Boulware, Israel-Hartle-Hawking and Unruh vacua. They mainly impose conditions on the components of energy-momentum tensor. Therefore we first provide the expressions of the components of stress-tensor (\ref{EMTPhi}). These will be useful for our main discussion as well.

The components of $T_{ab}$ in the background (\ref{rtmet}) in original coordinates are given by 
\begin{align}
    T_{tt} &= \f{c_w}{4}\left\{16p^2 + z^2 + 4ff^{\prime \prime} - (f^{\prime})^2\right\}~; \label{Ttt}\\ 
    T_{rr} &= \f{c_w}{4}\left\{\f{16p^2 + z^2 - (f^{\prime})^2}{f^2}\right\}~; \label{Trr}\\
    T_{rt} &= -2c_w \f{zp}{f}~. \label{Trt}
\end{align}
In null-null coordinates the components are
\begin{align}
    T_{uu} &=  \f{c_w}{4}\left[f~f^{\prime \prime} -  \f{(f^{\prime})^2}{2}\right] + \f{c_w \left(z+4p\right)^2}{8}~; \label{Tuuf}\\ 
    T_{vv} &= \f{c_w}{4}\left[f~f^{\prime \prime} -  \f{(f^{\prime})^2}{2}\right] + \f{c_w \left(z-4p\right)^2}{8}~. \label{Tvvf}
\end{align}
The integration constants $z,p$ are so far undetermined. However, suitably choosing these constants corresponds to different vacuum states for the quantum field $\Phi$ \cite{Banerjee_2009,Balbinot_1999}. This can be seen from the behaviour of the elements of the energy momentum tensor (\ref{EMTPhi}). We consider three important vacua - the Unruh vacuum \cite{PhysRevD.14.870}, the Israel-Hartle-Hawking vacuum \cite{israel1976thermo, PhysRevD.13.2188} and the Boulware vacuum \cite{PhysRevD.11.1404} (a summary of the definitions for these vacua are available in \cite{Balbinot_1999} as well). We describe these vacua through the components of the energy momentum tensor in Kruskal coordinates $U = -\frac{1}{\kappa_H}e^{-\kappa_H u},V = \frac{1}{\kappa_H}e^{\kappa_H v}$, where $\kappa_H$ is the surface gravity.

{\it Boulware vacuum: --} The Boulware vacuum \cite{PhysRevD.11.1404} is such that both \textit{in} and \textit{out} modes are of positive frequency with respect to $\p_t$. This state is asymptotically similar to the Minkowski vacuum. In this state, we have that both $T_{UU}$ and $T_{VV}$ are regular at infinity and hence, $T_{uu}, T_{vv} \to 0$ as $r \to \infty$. Thus, we have: 
\begin{align}
\begin{rcases}
z - 4p &= 0 \\  
z + 4p &= 0
\end{rcases}
\implies z = p = 0~.
\end{align}

{\it Israel-Hartle-Hawking vacuum: --} The Israel-Hartle-Hawking vacuum \cite{israel1976thermo, PhysRevD.13.2188} is a choice of vacuum state in thermal equilibrium. In this state, we have that both $T_{UU}$ and $T_{VV}$ are regular on the future horizon, and hence $T_{uu}, T_{vv} \to 0$ as $r \to r_H$. Denoting $\f{\p f}{\p r}\left. \right\vert_{r=r_H} = f'_H$, we have: 
\begin{align}
\begin{rcases}
z-4p &= \pm f^{\prime}_H \\ 
z+4p &= \pm f^{\prime}_H
\end{rcases}
\implies z = f^{\prime}_H \,\,\,\ {\textrm{and}}~~ p = 0~.
\end{align}

{\it Unruh Vacuum: --} The Unruh vacuum \cite{PhysRevD.14.870} is a choice of vacuum state which is useful to describe black hole evaporation. The state is chosen such that \textit{in} modes are chosen to be positive with respect to the timelike Killing vector $\p_t$ in Schwarzschild spacetime. In this state, we have that $T_{UU}$ is regular on the future horizon, and $T_{VV}$ is regular at infinity. This corresponds to demanding that $T_{uu} \to 0$ as $r \to r_H$ and $T_{vv} \to 0$ as $r \to \infty$. Thus, we have: 
\begin{align}
\begin{rcases}
z+4p &= \pm f^{\prime}_H  \\ 
z-4p &= 0
\end{rcases}
\implies z = \pm f^{\prime}_H/2 \,\,\,\,\ {\textrm{and}}~~ p = \pm f^{\prime}_H/8~. 
\label{B5}
\end{align}

We now see that discarding $p = 0$ naturally only leaves the Unruh vacuum as a possible choice of vacuum for out current construction of $\Phi$ fluid. Thus, we can compute the thermodynamic quantities of the anomalous $\Phi$ fluid in this state at different spatial limits. Further, in the following section we will show that only the upper set of signs for $z,p$ in the Unruh vacuum is physically meaningful to obtain well-defined fluid thermodynamic quantities at the horizon.

\section{Fluid parameters in Unruh vacuum}\label{UnruhFluid}
The previous discussion showed that for our present fluid construction, the Unruh vacuum is a suitable one. In this section we therefore will calculate the fluid parameters (\ref{B3}) -- (\ref{B4}) under this vacuum, particularly at two boundaries of the spacetime -- at the horizon $(r=r_H)$ and at asymptotic infinity $(r\to\infty)$. Therefore we will choose the values of the constants as given in (\ref{B5}). Although the expressions for components (\ref{Ttt}) -- (\ref{Trt}) are available in literature (e.g. see \cite{Balbinot_1999}), for the shake of our future discussion here we will again give them at these two extreme boundaries as well. For simplicity we choose the metric coefficient as that of a two dimensional Schwarzschild spacetime; i.e. $f(r) = 1- 2M/r$, where $M$ is the mass of the black hole.
\subsection{On the horizon}\label{UFHorizon}
The stress-tensor components are given by
\begin{align}
    T_{tt} \to -\f{c_w}{8}(f'_H)^2 \to -\f{1}{768M^2\pi}~; \,\,\,\
    T_{rr} \to -\infty~; \,\,\,\
    T_{rt} \to - \infty~, 
\label{B6}    
\end{align}
whereas the fluid parameters are
\begin{align}
    \rho^H &\to  \f{c_w}{8}\left[\f{5(f''_H)^2 - f'_Hf'''_H}{f''_H}\right] \to - \f{7}{768 M^2 \pi}~; \nonumber \\ 
    p_{iso}^H &\to -\f{c_w}{8}\left[\f{3(f''_H)^2 + f'_H~f'''_H}{f''_H}\right] \to \f{3}{256M^2\pi}~; \nonumber \\ 
    \kappa^H &\to \sqrt{-f''_H} \to \f{1}{\sqrt{2}M}~; \nonumber\\
    \eta^H &\to \f{c_w}{2}\sqrt{-f''_H} \to \f{1}{48\sqrt{2}M~\pi}~; \nonumber\\ 
    \theta^H &\to \f{1}{\sqrt{-f''_H}}\left[-f''_H + \f{(f''_H)^2+f'_H~f'''_H}{8f''_H} \right] \to \f{11}{16\sqrt{2}M}~; \nonumber \\ 
    \sqrt{q_aq^a} &\to \f{5}{1536M^2\pi}~.
\label{B7}    
\end{align}

The behaviour of the components (\ref{B6}) is exactly same as obtained in \cite{Balbinot_1999}. In the above $c_w=1/(24\pi)$ has been used.
We observe a few features of interest for the fluid parameters (\ref{B7}). Firstly, note that the energy density $\rho^H$ at the horizon is negative. Negative energy densities in QFT in the context of the Casimir effect and black hole near horizon are not new \cite{Freivogel_2014, FORD_2010}. The negative energy density obtained can be explained as follows. Computing the horizon energy density in the Unruh vacuum in Schwarzschild spacetime implicitly means we are computing the energy density near the horizon, relative to a known vacuum state. In our case, for ingoing freely falling observers the comparison is with the Minkowski vacuum. We can go one step further and show that the region of negative energy density outside the horizon, must be compensated by positive energy density behind the horizon. Firstly, from (\ref{B3}), we can calculate a general expression for the energy density in the Unruh vacuum in Schwarzschild spacetime given by,
\begin{align}\label{rhoout}
    \rho_{\text{out}} &= -\frac{4 M^3+4 M^2 r+2 M r^2+r^3}{192 \pi  M^2 r^3+96 \pi  M r^4}~.
\end{align}
This expression is valid for $r > r_H$, and hence we see that the total energy outside the horizon can be obtained as 
\begin{align}
    E_{\text{out}} &= \int_{r_H}^\infty~dV~\rho_{\text{out}} < 0~,
\end{align}
where $dV$ is the volume element for our 2D geometry. However, the total energy of the field over the full spacetime manifold must be positive, and thus the region behind the horizon must be a region of positive energy density compensating for the negative energy density outside, particularly near the horizon. This is also consistent with what is obtained in \cite{Freivogel_2014}. We also point out that the component $T_{tt}$ is negative, as also shown in \cite{Balbinot_1999}. We can now comment on the negative temperature obtained in Section \ref{thermofluidpar} for the solution $p \neq 0$ and $\kappa = \sqrt{\f{16p^2 - (z-f^{\prime})^2}{f}}$. As we have seen, this solution corresponds to the Unruh state where the energy density for the fluid remains negative for any finite $r > r_H$ (see (\ref{rhoout})), increasing until it asymptotes to $0$ at infinity.  Consider an ingoing freely falling observer. As the observer moves towards the horizon, the energy density $\rho_{\text{out}}$ in (\ref{rhoout}) becomes more negative i.e. decreases and hence the change $\delta\rho < 0$. From the first law of thermodynamics we must have $\delta\rho = T\delta s$ (where $s$ is the entropy density). A negative temperature is consistent with a positive change in entropy, and hence $\delta s>0$, as required by the second law. Finally, note that the normalisation $N_0$, and hence $\kappa$ vanish at spatial infinity. Thus, the inversion used in (\ref{kappaT}) to obtain $T$ is no longer valid and hence the temperature at infinity is no longer given by $-c_w$. This is again consistent with the fact that at infinity the energy density vanishes. Ultimately, in our interpretation all thermodynamic quantities arise due to the anomaly and since the anomaly vanishes at infinity, this is expected. Importantly, we see that dissipative quantities such as $\kappa, \eta$ and the norm $q_aq^a$ have finite values at the horizon. These dissipative effects arise through the higher derivatives in $T_{ab}$ are ultimately due to the anomaly, which contributes significantly near the horizon. The above limits have been calculated with $z = 4p = f'_H/2$ (i.e. the upper signs in (\ref{B5})). Using the negative signs gives divergent results for $\rho, p_{\text{iso}}$ and $\theta$ is no longer real. Hence, we discard this solution for the fluid description through Eckart's formalism. 

Let us end this subsection with more discussion on the concept of negative temperature. According to the ordinary definition of temperature $(T)$, the entropy $(S)$ should decrease with the increase in internal energy $(U)$ for systems with negative temperature. But we would like to stress that there is no mathematical objection to consider the decrease of $S$ with the increase in $U$ to establish many thermodynamic theorems (see \cite{PhysRev.103.20} for an extensive discussion on this statement). Although such systems are not generally realized in practice, but they may be theoretically devised, as well as closely experimentally realized also. Statistical mechanics shows that if a system has upper bound in energy, then it can acquire negative $T$. In such situation when all the constituents are in lowest energy state or in highest energy state, the entropy becomes zero. However, when these are distributed among various states, then the entropy of the system is large. Therefore, one can say that the entropy $S$ starts from zero in the lowest energy state and then increase with the increase of $U$ and finally again vanishes for some other value of $U$. It has been argued in \cite{PhysRev.103.20}, that the thermodynamic theorems apply equally well for systems with $-T$. However, certain thermodynamic quantities are needed to be watched out in that case, particularly ``heat'' and ``work''. For $-T$ system, the sign of these quantities should be reversed. If two systems, respectively with positive and negative temperatures, are in contact, then interestingly the heat flows from negative $T$ system to positive $T$ system. In that sense $-T$ system is ``hotter'' than the $+T$ system, although algebraically it is the other way around. Therefore, the $-T$ system is ejecting heat. Hence, from the perspective of the $-T$ system, heat exchange is negative and hence the second law of thermodynamics is preserved. As discussed above, we find our present situation to be similar. We saw that $\delta \rho<0$, and therefore we can accommodate negative $T$. The theoretical idea of $-T$ has also been observed in experiments for few specific systems, like nuclear spins, putting the atoms into optical lattice (created by laser beams), {\it etc.} which have upper bounds in energy (see also a recent review \cite{Baldovin_2021} on the negative temperature and corresponding interpretations). However it must be pointed out that the $-T$ is observed only with respect to some of the degrees of freedom of a full system; e.g. in spin system only with respect to spin. Consideration of all degrees of freedom (like molecular vibrational, electronic etc. degrees of freedom of the full system) will lead to sensible positive $T$ for the total system. Similarly, here we are considering only the anomalous part for the fluid, not all the constituents in the energy-momentum (EM) tensor (like non-anomalous EM, which is the standard non-gravitational part). It is expected that inclusion of all possibilities will lead to positive temperature. Since our attention is only to investigate the gravitational anomaly part, the appearance of $-T$ is not unexpected in this context. In addition our result $T=-c_w$ depends on the crucial choice (\ref{spatialder}). This, according to our opinion, is the simplest one to provide an analytical solution of it. Such a choice has also been considered earlier as well \cite{Faraoni:2021jri, Faraoni:2021lfc, Faraoni:2022gry}. However, more general situation probably may improve the results, more particularly we may get a positive temperature. Since this generalization provides more complicity in the analytical treatment of the equations we chose the simplest one for the moment to shed some light in this direction. But it would be interesting to investigate a generalization of such choice.

\subsection{At Infinity}\label{UFInf}
In the asymptotic infinity region, the components of energy-momentum tensor are
\begin{align}
    T_{tt} &\to \f{c_w}{8}(f'_H)^2 \to \f{1}{768M^2\pi} = \f{\kappa_H^2}{48\pi}~; \,\,\,\
    T_{rr} \to \f{c_w}{8}(f'_H)^2 \to \f{1}{768M^2\pi} = \f{\kappa_H^2}{48\pi}~;\nonumber \\ 
    T_{rt} &\to -\f{c_w}{8}(f'_H)^2 \to -\f{1}{768M^2\pi} = -\f{\kappa_H^2}{48\pi}~,
\label{B8}
\end{align}
where $\kappa_H$ is the surface gravity of the black hole.
These are as expected (see \cite{Balbinot_1999}). Most importantly, it may be mentioned that $T_{uu} (r\to\infty)$ gives us the Hawking flux $\kappa_H^2/48\pi$ as seen by the asymptotic observer. This is well known that the Unruh vacuum is suitable one to discuss Hawking effect and then through anomaly approaches one calculates the correct Hawking flux \cite{Balbinot_1999,Banerjee_2009,Banerjee_2009_02} under the use of (\ref{B5}) in (\ref{Tuuf}) along with $r\to\infty$ limit.

Now we are interested to investigate the behaviour of the fluid parameters at $r\to\infty$. These turn out to be as follows:
\begin{equation}
    \rho^\infty \to  0~; \,\,\ 
    p_{iso}^\infty \to 0~; \,\,\
    \kappa^\infty \to 0~; \,\,\,
    \eta^\infty \to 0~; \,\,\
    \theta^\infty \to 0~; \,\,\
    q_aq^a \to 0~.
\label{B9}    
\end{equation}
Note that the normalisation of the fluid velocity $N_0$ vanishes at infinity in the Unruh vacuum (see (\ref{N0})). However, this is not a problem since the entire fluid description arises from the anomaly only, which vanishes at infinity. Interestingly, all the parameters vanish at asymptotic infinity. This is consistent with the fact that our energy-momentum tensor is completely contributed by the anomaly in the trace which vanishes at infinity.

Before proceeding further, let us discuss very important points related to diffusion flux $\nu_a$ and the negative value of fluid temperature. The presence of $\nu_a$ confirms that rather than $nu^a$, the number current $N^a = nu^a+\nu^a$ is conserved (i.e. $\nabla_aN^a=0$), where $u^a\nu_a=0$ and $n=-N^au_a$. Based on the presence of such diffusion, a generalised version of Eckart's formalism has been developed in \cite{Andersson:2006nr}, where $\nu^a$ is given by (\ref{nu}). This usually happens due to the presence of potential term in the action (see discussion around Eq. (1.15) of \cite{Faraoni:2022gry}). In this case, as suggested in \cite{Faraoni:2022gry}, the chemical potential $\mu$ will be a relevant quantity to introduce, the presence of which is generated by non-vanishing nature of $\nu_a$. Usually the conservation of $nu^a$ is the consequence of the shift symmetry of the system (see e.g. \cite{Dubovsky:2011sj}). Particularly, the Noether current corresponding to this symmetry is identified as this quantity. However, in presence of certain terms like potential in action, the symmetry is lost and one encounters diffusion type situation. For our action (\ref{PRDV41}) the situation is same. Due to the term $R\Phi$, the symmetry is broken and therefore $nu^a = \frac{\partial S_P^{(g)}}{\partial (\nabla_a\Phi)} = \frac{c_w}{2}\nabla^a\Phi$ is not conserved. However one can identify $\nu^a$ through $\nabla_a(nu^a) = \frac{c_w}{2}\Box\Phi = \frac{c_w}{2} R$, where in the last step (\ref{phieom}) has been used. Then $\nabla_aN^a=0$ implies $\nabla_a\nu^a = - \frac{c_w}{2} R$. Now for our metric (\ref{rtmet}), we have $R = -f''$ and since the background is independent of time one has fluid quantities which are time independent. Then we find $\nabla_a\nu^a = \partial_r\nu^r = \frac{c_w}{2}f''$ where $r$ represents only the radial component and so the radial component is given by $\nu^r = \frac{c_w}{2} f' + C$, where $C$ is a constant. If we demand that $\nu^r$ must vanish at $r\to\infty$, then one needs to choose $C=0$. The time component can be determined by $u^a\nu_a=0$, which gives $\nu^t = \frac{u^r\nu^r}{f^2 u^t}$, where the time and radial components of fluid velocity is determined by (\ref{4vel}). Using these values of $\nu^a$, the chemical potential $\mu$ can be determined by (\ref{nu}). In fact, contracting (\ref{nu}) with $\nu_a$ one obtains
\begin{equation}
\mu  = \int dr \frac{\nu^a\nu_a}{\sigma c_w\nu^r}~.
\end{equation}
The above has been found by using the fact that $\mu$ is independent of time for our background metric and $T=-c_w$.

Since temperature appears to be negative, there is a concern of the validity of the second law of thermodynamics. Remember that the present analysis is based in Eckart's formalism. In this case the local entropy current satisfies the following relation
\begin{equation}
\nabla_as^a = \frac{q^aq_a}{\kappa T^2} + \frac{\nu^a\nu_a}{\sigma T^2} + \frac{p_{\textrm{vis}}^2}{\zeta T} + \frac{\pi^{ab}\pi_{ab}}{2\eta T}~. 
\end{equation}
The explicit derivation is given in \cite{Andersson:2006nr} (see Eq. (279)).  For (extended) Eckart's formalism one sets  $p_{\textrm{vis}} = 0$ and then the above reduces to
\begin{equation}
\nabla_as^a = \frac{q^aq_a}{\kappa T^2}  +  \frac{\nu^a\nu_a}{\sigma T^2}+ \frac{\pi^{ab}\pi_{ab}}{2\eta T}~. 
\end{equation}
Now as $q_a$ and $\nu^a$ are spacelike, the first and second terms must satisfy $\frac{q^aq_a}{\kappa T^2}\geq 0$ and $\frac{\nu^a\nu_a}{\sigma T^2}\geq 0$, respectivey. However since in our case $T$ is a negative quantity, the second term can be a negative one. But, although the present $\pi_{ab}$ is non-vanishing, one can demonstrate that $\pi^{ab}\pi_{ab} = 0$. This we discuss below.

In this analysis, we consider a four-velocity of the form  (\ref{4vel}). For such a (timelike) congruence, the shear tensor $\sigma_{ab}$ is defined to be $\sigma_{ab} = \nabla_bu_a + \dot{u}_au_b - \theta h_{ab}$ where $\theta$ is the expansion scalar, $\dot{u}_{a}$ is the acceleration and $h_{ab} = g_{ab} + u_au_b$ is the projection tensor. The values of these expressions can be found from equations (\ref{rhoed}) -- (\ref{4accln}). Using these, $\sigma_{ab}$ is computed to be, 
\begin{equation}
    \sigma_{ab} = \frac{1}{\sqrt{-N_0}}\left\{\nabla_a\nabla_b\Phi - \frac{Z_{ab}}{N_0} - g_{ab}\left(R - \frac{M_0}{N_0}\right) + \nabla_a\Phi\nabla_b\Phi \frac{R}{N_0}\right\}~,
\end{equation}
where we use the notation  $N_0 = \nabla_i\Phi\nabla^i\Phi$, $M_0 = \nabla^a\Phi\nabla^b\Phi\nabla_a\nabla_b\Phi$ and $Z_{ab} = \nabla_a\Phi\nabla^i\Phi\nabla_{b}\nabla_i\Phi + \nabla_{b}\Phi\nabla^i\Phi\nabla_{a}\nabla_i\Phi$.
Now using the above equation in $\pi_{ab} = - 2 \eta \sigma_{ab}$ and substituting the value of $\eta$, given by Eq. (\ref{B2}), i.e. $\eta = (c_w/2)\sqrt{-N_0}$, one finds the viscous stress as
\begin{eqnarray}
\label{piabeqn}
    \pi_{ab} = c_w\left\{g_{ab}\left(R - \frac{M_0}{N_0}\right) - \nabla_a\nabla_b\Phi - \nabla_a\Phi\nabla_b\Phi \frac{R}{N_0} + \frac{1}{N_0}Z_{ab}\right\}~.
\end{eqnarray}
Next check that $\pi_{ab}u^a = 0 =\pi_{ab}u^b$ and $\pi_{ab}h^{ab} = \pi_{ab}g^{ab} = 0$ as required. Employing these we calculate the product $\pi_{ab}\pi^{ab}$ to be,
\begin{eqnarray}
    \pi_{ab}\pi^{ab} &= c_w^2\left[R^2 - 2 R \frac{M_0}{N_0} - 2 \left(R - \frac{M_0}{N_0}\right)^2 - 2 \frac{\nabla_a\nabla_b\Phi Z^{ab}}{N_0} + \frac{Z_{ab}Z^{ab}}{\left(N_0\right)^2} + \nabla_a\nabla_b\Phi~\nabla^a\nabla^b\Phi\right]~.
\end{eqnarray}
Our present fluid description is in $(1+1)$-dimensions and the relevant metric we considered to be Eq. (\ref{rtmet}). In this metric $\Phi$ is given by Eq. (\ref{PRDV31}). Substitution of this and the required quantities for the metric it is easy to verify that
\begin{equation}
    \pi_{ab}\pi^{ab} = 0~.
\end{equation}
This confirms that the local version of entropy increasing theorem is satisfied for these identified fluid parameters.

Finally, we mention various important aspects of the present fluid description. Earlier attempts revealed that the anomalous correction provides second order correction to fluid constitutive relations. 
Here, we investigate whether the anomalous part can be included at the first order theory or not. In doing so, contrary to earlier investigations \cite{Landsteiner:2011cp,Dubovsky:2011sk,Valle:2012em,Jensen_2013,Banerjee:2013qha,Banerjee_2014,Majhi_2014,Majhi:2014hsa,Banerjee:2014cya,Banerjee:2014ita}, we consider the anomalous fluid as a scalar fluid and the fluid 4-velocity is being defined through auxiliary scalar field which defines the anomalous stress-tensor. The fluid structure has been investigated in detail till this section. A general construction of fluid description of anomalous stress-tensor is provided in Section 4. However, to get a better understanding of this description, we prefer to choose an investigation of the fluid parameters and its thermodynamic quantities on a static background. This is done in sections 4, 5 and in the present one. We observed that the Unruh vacuum is the suitable one, which is in contrary to earlier description where Israel-Hartle-Hawking vacuum \cite{Majhi_2014,Majhi:2014hsa} was found to be the relevant one. So, our main motivation of the present study is to provide a first order fluid description in presence of gravitational anomaly. It may be mentioned here that, there are several benefits to incorporate any correction to fluid constitutive relations at the first order in fluid parameters. The fluid equations will be less non-linear compared to higher order analysis. Therefore, unique solution can be obtained with less number of boundary conditions and hence obtaining a solution will be less difficult. Consequently, the numerical analysis with these equations becomes feasible. Also, higher order theory requires extended space of variables and so in order to desire to work with less number of variables one must design a first order theory. Moreover, the thermodynamics of the fluid within first order formalism is less cumbersome.

To conclude the discussion, we mention that in $(1+1)$ spacetime dimensions the Cardy formula relates the pressure of conformal fields with the left and right handed central charges through the relation $p_{(\text{Cardy})} = 2\pi T_0^2 (c_L+c_R)/24 = 4\pi^2 c_w T_0^2$, where $T_0 = \kappa_H/(2\pi)$ is the Hawking temperature in this case \cite{Jensen_2013}. For Schwarzschild black hole we have $p_{(\text{Cardy})} = 1/(384\pi M^2)$. Interestingly this is equal to $\rho^H + p^H_{iso}$ of our described fluid (see Eq. (\ref{B7})). However this similarity is purely coincidence, and has no physical logic behind this.

\section{Connection with anomaly cancellation technique for Hawking flux} \label{Hawking}
So far the discussion for our anomalous fluid through Eckart's formalism revels that the most suitable choice of vacuum should be Unruh vacuum.
Also we mentioned that the Unruh vacuum condition on the anomalous stress-tensor provides the required value of Hawking flux. Since this choice of vacuum is relevant for both calculation of Hawking flux through anomaly as well as the present anomalous fluid, then it is expected that there might be a relation between the anomaly cancellation approach and Eckart's formalism for the anomalous fluid. Particularly, it might be a fact that the Hawking flux can be obtained from $T_{uu}$ component by imposing suitable conditions on the fluid parameters (\ref{B3}) -- (\ref{B4}), which are equivalent to Unruh vacuum condition. Then these will be an alternative interpretation of this particular choice of vacuum. Let us now proceed toward this goal.

Note that in the asymptotic region $r\to\infty$, the right hand side of (\ref{Trace}) vanishes and hence it is anomaly free. So one expects that the fluid parameters, like $\eta$ and $\kappa$ must vanish as the fluid energy-momentum tensor is completely determined by the gravitational trace anomaly. Then (\ref{B10}) or (\ref{B11}) implies that $z=\pm 4p$. On the other hand since near the horizon we have anomaly, it is then expected that $\eta$ or $\kappa$ must not vanish; rather they must be finite. Now use of $z=\pm 4p$ in (\ref{B10}) yields
\begin{equation}
\kappa = \sqrt{\frac{-(f'^2\mp 8pf')}{f}}~.
\label{B12}
\end{equation}
This will be finite at the horizon provided we choose $p=\pm f'_H/8$. Same can be yielded from finiteness of $\eta$ on the horizon as well. Therefore we have two sets of choices: either $z=4p$ with $p=f'_H/8$ or $z=-4p$ with $p = - f'_H/8$. Note that these choices are coming within the fluid description. Right now these do not have anything to do with choices by imposition of different vacuum conditions. However we will show below that among these two choices one is relevant and then can be related to Unruh condition.

For the first choice $T^r_t = -2c_w zp$ becomes negative whereas the other choice leads to positive value of the flux. Therefore we choose $z=4p$ with $p = f'_H/8$ to get positive value of the flux. Note that such a choice is consistent with Unruh vacuum condition (\ref{B5}) and then one obtains the correct expression for Hawking flux $T_{uu} (r\to\infty)=\frac{\kappa_H^2}{48\pi}$. Moreover we saw that both Boulware and Israel-Hartle-Hawking vacua are consistent with the choice $p=0$. However, as we have seen earlier, such choice in fluid description leads to various issues. The definition of four-velocity (\ref{4vel}) is then no longer well defined. The normalization factor in $u^a$ looses its sense and $u^a$ becomes spacelike. This is because for the present metric we have
\begin{equation}
u^t = \frac{4p}{\sqrt{16p^2f - f(z-f')^2}}~,
\label{S5}
\end{equation}  
and then $u^t$ vanishes for $p=0$. This implies that these two vacua are not well posed with the present fluid description.
Hence the imposition of relevant conditions on the fluid parameters can also provide exactly identical description of Unruh vacuum.

Another important observation is as follows. $u^t$ component is given in (\ref{S5}) and $u^r$ is 
\begin{equation}
u^r = \frac{\sqrt{f}(z-f')}{\sqrt{16p^2 - (z-f')^2}}~.
\label{RV1}
\end{equation}
Note that the denominators for both the components vanish at $r\to\infty$ for the choice $z=4p$. Therefore it may appear that the approach suffers from singularity at $r\to \infty$. However we will show that such is not true. The components of the velocity in Kruskal coordinates ($U = -(1/\kappa_H) e^{-\kappa_H u}$; $V=(1/\kappa_H) e^{\kappa_H v}$) are given by
\begin{equation}
u_U = -\frac{u_u}{\kappa_H U}; \,\,\,\,\ u_V = \frac{u_v}{\kappa_H V}~.
\label{RV2}
\end{equation}
Now in Unruh vacuum the positive frequency {\it in} modes are defined with respect to the Schwarzschild timelike Killing vector; while that for {\it out} modes are identified through the Kruskal $U$ coordinate. Therefore the vacuum must be Minkowskian at $r\to\infty$ so that there is no ingoing quantity. Hence one must have $u_v\to 0$ at $r\to \infty$. On the other hand $u_U$ must be regular at $r=r_H$. However, at the horizon we have $U=0$ and so one must has $u_u=0$ at $r=r_H$ to make $u_U$ regular (see \cite{Balbinot_1999,Banerjee_2009,Majhi:2014hsa} for a detailed discussion on these conditions). For the present analysis the components (\ref{S5}) and (\ref{RV1}) of $u^a$ in null-null coordinates ($u,v$) are given by
\begin{equation}
u_u = -\frac{\sqrt{f}}{2}\sqrt{\frac{4p + (z-f')}{4p-(z-f')}}~; \,\,\,\,\ u_v =  -\frac{\sqrt{f}}{2}\sqrt{\frac{4p - (z-f')}{4p+(z-f')}}~.
\label{RV3}
\end{equation}
These for $z=4p = f'_H/2$ reduces to
\begin{eqnarray}
&&u_u = -\frac{\sqrt{f}}{2}\sqrt{\frac{8p -f'}{f'}} = -\frac{\sqrt{f}}{2}\sqrt{\frac{f'_H -f'}{f'}}~; 
\nonumber
\\
&&u_v =  -\frac{\sqrt{f}}{2}\sqrt{\frac{f'}{8p-f'}} = -\frac{\sqrt{f}}{2}\sqrt{\frac{f'}{f'_H-f'}}~.
\label{RV4}
\end{eqnarray}
Observe that the above components are compatible with the Unruh conditions; i.e. $u_u\to0$ for $r\to r_H$ and $u_v\to 0$ for the limit $r\to\infty$. Now in Kruskal coordinates these take the forms:
\begin{eqnarray}
u_U =  \frac{e^{\kappa_Hu}}{2}\sqrt{\frac{f(f'_H -f')}{f'}}~; \,\,\,\ u_V = -\frac{e^{-\kappa_H v}}{2}\sqrt{\frac{ff'}{f'_H-f'}}~.
\label{RV5}
\end{eqnarray}
Note that 
$r\to\infty$ implies $u\to -\infty$ and $v\to \infty$. Therefore in the limit $r\to\infty$ the exponential factor in $u_U$ converges to zero rapidly than the divergence of its term within the square root, which makes $u_U$ finite. Similarly $u_V$ approaches to zero for $r\to\infty$. Therefore at the asymptotic infinity region all the components of $u_a$ in Kruskal coordinates have proper behavior. Since Kruskal coordinates are more appropriate to Unruh vacuum, the well behaved components in these coordinates make our analysis singularity free at $r\to\infty$. Therefore imposition of the condition on the fluid parameters at infinity and connecting it with the Unruh vacuum as well as Hawking flux is well structured within our choices.
Hence it seems that the Eckart's formulation of thermodynamics of the anomalous fluid is equally capable of calculating Hawking flux through anomaly cancellation technique where imposition of Unruh vacuum can be interpreted as demanding the vanishing of viscosity and thermal conductivity of our fluid at infinity.

Before concluding the section let us mention few comments on this similarity between the Unruh condition ($T_{uu} \to 0$ at $r\to r_H$ and $T_{vv}\to 0$ at $r\to\infty$) and imposition of condition $\eta$, $\kappa$ $\to 0$ at $r\to\infty$ either of which is required to find the Hawking flux in anomaly approach.
\begin{itemize}
\item Our main motivation in this analysis is to provide a first order description of anomalous fluid, contrary to earlier description within the second order formalism. In doing so, we have found a connection between the valid fluid interpretation and the required condition to obtain Hawking flux in anomaly cancellation approach. Therefore, the above discussion can be considered as a by-product, rather than the main goal of the present work.  Moreover, such a similarity seems to be a completely technical point, rather than providing any concrete physical picture towards deeper understanding of the gravitational phenomenon. Hence this section can be consider as a side discussion. 
\item It may be mentioned that the present stress-tensor is that corresponding to the external matter lying on the two dimensional black hole background. More precisely, the scalar fields at the quantum level do not respect the conformal symmetry and therefore upon quantization the trace of the energy-momentum tensor becomes non-vanishing. Interestingly, the non-vanishing trace is given by the Ricci scalar of the background geometry. In this analysis, as mentioned earlier, we targeted to interpret this as the corrections to fluid constitutive relations due to the presence of gravity. However contrary to the previous attempts to incorporate it as second order corrections, we liked to have a first order formalism. Since this stress-tensor is for the external matter, therefore we do not expect to provide the properties of black hole through the identified fluid parameters. Hence such matter properties (which a different system than the black hole) should not have any connection with the gravitational degrees of freedom. Further we stress that in earlier description (the second order formalism) the black hole temperature (the Tolmann expression) has been taken as the temperature of the anomalous fluid. However we found here that the first order formalism is suitable with a constant temperature. 
So the underlying microscopic description of thermodynamics of the black hole cannot be illuminated through studying the external matter, unless the quantum description of black hole is taken into account.
Furthermore, the Hawking flux is calculated by the observer is at asymptotic infinity where, as we mentioned, the fluid description ceases to exist because of the vanishing of anomaly. Therefore direct relation between the fluid quantities and Hawking flux does not seem to be apparent. 
\item As we pointed out just now, the information which we are using is the behavior of quantum fields under gravity. So gravity is not directly analyzed and hence complete information about the microscopic structure of gravity cannot be illuminated in this way. Moreover the fields are quantized outside the black hole horizon. Therefore we do not expect the present analysis is capable of describing the microscopic degrees of freedom behind the horizon (known as islands) responsible for black hole entropy. Similarly it is very hard to find its any connection in the direction of singularity resolution.
\end{itemize}

\section{Conclusions} \label{conclusion}
The anomalous fluid within Eckart's formalism \cite{PhysRev.58.919} (up to a generalization provided in \cite{Andersson:2006nr}) has been discussed. Since in this formulation the covariant conservation of EMT is necessary, we consider two spacetime dimensional situation where only trace anomaly presents due to gravity. This also makes Eckart's approach a more reasonable approximation to study relativistic thermodynamics since the causality and instability issues that may normally arise are less likely to affect our (1+1) dimensional system. Thus, for a first analysis we work with Eckart's formalism and still obtain results consistent with existing literature such as \cite{Balbinot_1999}. Using more sophisticated models such as those due to Israel and Stewart \cite{israel1979transient}, in the future may provide a clearer picture and this is one of the possible future directions. Like the other applications of Eckart's formalism in different gravitational theories \cite{Faraoni:2021lfc,Faraoni:2021jri,Giardino:2022sdv,Faraoni:2022jyd,Faraoni:2022doe,Faraoni:2022fxo,Faraoni:2022gry,Faraoni:2023ann,Faraoni:2023hwu,https://doi.org/10.48550/arxiv.2302.00373,Giusti:2021sku,Miranda:2022wkz,Faraoni:2023hwu}, here we define the four-velocity of the fluid as covariant derivative of the auxiliary scalar field which describes the local form of the effective action, namely the Polyakov action \cite{polyakov1981quantum}. Only the anomalous part of the EMT of the fluid has been considered. In static spacetime with Killing horizon, we observed that the fluid parameters are finite and non-vanishing at the horizon; whereas they vanish at the asymptotic spatial infinity. In this discussion we found that the setup is well described only in Unruh vacuum. 

The present approach is clearly based on the crucial conservation relations like $\nabla_a T^{ab} = 0$ and $\nabla_aN^a = 0$. Hence the existence of $\mu$, is well motivated and related to the conservation of the current $N_a = nu_a+\nu_a$. Furthermore it may be mentioned the conservation of $nu^a$ can be violated when there is a source or sink. This is the case when the EMT contains a potential term and therefore the shift symmetry is broken. This exactly happens in our case as well. However one can define a total $N_a =nu_a+\nu_a$ which is conserved. In this regard it may be mentioned that, as far as we know whether such a conservation is a consequence of any particular symmetry of the theory, is not clear in literature. 
In summary, the analysis is well within the general idea of usual hydrodynamics. However in usual hydrodynamics the velocity parameter is a given quantity and the EMT is expressed in terms of $u^a$. But since the present EMT is not in this form, one needs to first identify $u^a$ to map it with the usual form of EMT.  As there exists no unique way to identify it, we followed a particular prescription given in literature. However such a choice is different from other works and hence the results are valid only within this parametrization.

Having the relevance of Unruh vacuum in Eckart's way of studying anomalous fluid, we provide a possible connection with the calculation of Hawking flux through anomaly approach \cite{Christensen:1977jc,Balbinot_1999,Robinson:2005pd,Iso:2006wa,Banerjee:2007qs,Banerjee:2007uc,Banerjee_2009,Banerjee_2009_02}. We argued that the relevant conditions on the fluid parameters can give a parallel description of Unruh vacuum. Particularly we mentioned that as the anomaly is non-vanishing at the horizon and it vanishes at infinity, the fluid parameters must be finite at the horizon while they must vanish at infinity. We found that such a demand is enough to produce the environment of Unruh vacuum and therefore one found correct value of Hawking flux through the relevant component of EMT.

We like to point out that the introduction of gravitational anomaly is on the basis of semi-classical approach. The background spacetime is taken to be classical one and the existing quantum fields are minimally coupled to gravity. In this case the fields are considered as external object and their quantum analysis is done on the given classical spacetime. Such an approach is well accepted as long as the back-reaction effect of the fields on the background spacetime is being neglected (see e.g. \cite{Book1}). Quantization of massless  scalar fields on a two-dimensional spacetime with horizon leads to breaking of conformal symmetry and thereby one has trace anomaly. This is manifested as the non-vanishing  trace of the energy-momentum tensor corresponding to the field at the quantum level which does not appear at the classical regime. In the present analysis we are not incorporating the back-reaction effect and therefore the background spacetime can be considered as (\ref{rtmet}). The same was also taken in various existing articles in literature, like \cite{Christensen:1977jc} -- \cite{Banerjee_2009_02} and  \cite{Valle:2012em} -- \cite{Banerjee:2014ita} (also can be followed various chapters in text book like \cite{Book1}). However incorporation of back-reaction effect in this analysis can be interesting which we leave for the future.  

Finally, we mentioned that the original Eckart's formalism requires $\nabla_a T^{ab}=0$ and so here we considered only the non-chiral theory. It would be interesting to extend this formalism for theories where the conservation violates. In this case the chiral cases can be considered, particularly for the energy-momentum tensor from Leutwyler action \cite{Leutwyler:1984nd}. Moreover it would be interesting to investigate the higher dimensional situations. Hope we will be able to illuminate these areas in future.

\vskip 3mm
{\bf Acknowledgement:}
The research of all the authors is supported by Science and Engineering Research Board (SERB), Department of Science $\&$ Technology (DST), Government of India, under the scheme Core Research Grant (File no. CRG/2020/000616). 

\vskip 5mm
\appendix

\section*{Appendices}

\section{Eckart's fluid: in a general framework}\label{BD1}
In general, the fluid may be imperfect which means that it contains viscous stresses and dissipative behaviour. One way to describe this is to consider the velocity gradient $\nabla_au_b$. We can decompose the velocity gradient as follows:
\begin{equation}\label{vdecomp}
    \nabla_au_b = \sigma_{ab} + \f{1}{3}\theta h_{ab} + \o_{ab} - \dot{u}_au_b~,
\end{equation} 
where 
$\sigma_{ab}$ is the symmetric traceless part, called the shear; $\theta = \nabla_au^a$ is the trace, called the expansion; and $\o_{ab}$ is the antisymmetric part, called the vorticity.
A non-zero velocity gradient can lead to viscosity and anisotropic stresses which modify the form of the stress tensor (which is no longer diagonal) as 
\begin{equation}
    \Pi_{ab} =  p_{\text{iso}}h_{ab} + \pi_{ab}~,
\end{equation}
where we considered $p_{\text{vis}} = 0$.
Here we have introduced the traceless symmetric tensor $\pi_{ab}$ which is the viscous stress tensor. Including the effect of a heat flux $q^a$, the energy momentum tensor of an imperfect fluid is given as 
\begin{equation}\label{imtab01}
    T_{ab} = \rho u_a u_b + \Pi_{ab} + q_au_b + q_bu_a~,
\end{equation}
which is Eq. (\ref{emfull}).
The heat flux $q^a$ is purely spatial by construction i.e. $u_aq^a = 0$.

We will now write down the fluid dynamical and thermodynamic relations for the fluid using Eckart's formalism. Using the mass density $m$ of a fluid, we define the mass current $m^a = mu^a$ and the conservation of mass takes the form: 
\begin{equation}
   \nabla_am^a = 0~.
\end{equation}
Thus, from the normalisation of the fluid velocity, $m = \sqrt{-m^am_a}$. We choose a stack of (spacelike) hypersurfaces orthogonal to the (timlike) fluid velocity. The induced metric on it is given by (\ref{indmet}). From the energy momentum tensor of the imperfect fluid (\ref{imtab01}), we can project out quantities that describe the fluid in the following way: 
\begin{enumerate}
    \item $w = T^{ab}u_au_b = \rho$,  the internal energy density,
    \item $w^a = - T^{bc}u_b\text{ }h^a_c = q^a$, the heat flux, 
    \item $w^{ab} = T_{pq}h^a_ph^b_q = \Pi^{ab}$, the stress tensor~.
\end{enumerate}
By construction, note that $u_aq^a = u_a\pi^{ab} = 0$. 
We want to now express the first law in this formalism. To begin with, we define the derivative with respect to the proper time using our four velocity as, $D = d/d\tau = u^a\nabla_a$. Now, because the energy momentum tensor (\ref{imtab01}) is conserved,
\begin{align}\label{fls1}
    \nabla_a(-u_bT^{ab}) + T^{ab}\nabla_au_b = 0~.
\end{align}
But, from (\ref{imtab01}) we have
\begin{eqnarray}
    &&\nabla_a\left(-u_bT^{ab}\right) = \nabla_a\left(q^a + \rho u^a\right) = mD\left(\f{\rho}{m}\right) + \nabla_aq^a~; 
    \label{fls2} 
    \\
     &&T^{ab}\nabla_au_b = \Pi^{ab}\nabla_au_b + q^bDu_b~. 
    \label{fls3}
\end{eqnarray}
Using (\ref{fls2}), (\ref{fls3}) in (\ref{fls1}), we obtain the first law of thermodynamics 
\begin{equation}
    mD\left(\f{\rho}{m}\right) + \left(\nabla_aq^a + q^aDu_a\right) + w^{ab}\nabla_au_b = 0.
\end{equation}
From $w^{ab}$, we define the viscous stress tensor as
\begin{equation}
    \pi_{ab} =  \Pi_{ab} - p_{\text{iso}}h_{ab} = w^{ab} - p_{\text{iso}}h_{ab}~.
\end{equation}
We assume that the viscous stress tensor is linear in the velocity gradients and proportional to the (scalar) coefficient of viscosity $\eta$. Consideration of further restrictions $u_a\pi^{ab} = \pi^a_a = 0 \implies p_{\text{iso}} = \Pi^a_a$ (in $(1+1)$ dimensions) 
suggests a form for the viscous stress tensor as 
\begin{equation}\label{piab}
    \pi_{ab} = - \eta \left\{h_{a p}h_{b q} \left(\nabla^{q}u^{p} + \nabla^{p}u^{q}\right) - {2}h_{ab}\nabla_{p}u^{p}\right\} = - 2 \eta \sigma_{ab}~. 
\end{equation}

We now wish to express the second law in Eckart's formalism. Let the particle number density in the absence of diffusion be $n$ and the number current is then $n^a = nu^a$. Consider the introduction of a diffusion flux $\nu^a$. Then, the total number current is given by
\begin{equation}\label{Na}
    N^a = n^a + \nu^a.
\end{equation}
By construction $u_a\nu^a = 0$ i.e. the diffusion flux $\nu^a$ is spatial, so that the number density can be obtained as $n = - u_aN^a$. The total number current is conserved i.e. $\nabla_aN^a = 0$, so we can define the fluid chemical potential $\mu$. Now, the entropy current in the system can be taken as
\begin{equation}
    s^a = su^a + \beta q^a - \lambda \nu^a~,
\end{equation}
where $s$ is the entropy density and $\beta,\lambda$ are unknown parameters. We can fix these parameters by imposing the second law $\nabla_as^a \geq 0$. The divergence is, 
\begin{equation}\label{sl1}
    \nabla_as^a = u^a\nabla_as + s\nabla_au^a + q^a\nabla_a\beta + \beta\nabla_aq^a - \lambda \nabla_a\nu^a - \nu^a\nabla_a\lambda~.
\end{equation}
Consider the first term which can be rewritten as, 
\begin{equation}
    u^a\nabla_a\left(n\f{s}{n}\right) = u^an\nabla_a\left(\f{s}{n}\right) + \left(\f{s}{n}\right)u^a\nabla_an~.
    \label{BD2}
\end{equation}
Now the usual first law of thermodynamics (per unit volume), 
\begin{align}
    &\nabla_a\rho = T\nabla_as + \mu \nabla_an = T \nabla_a\left(n \f{s}{n}\right) + \mu \nabla_an \nonumber~,
\end{align}
yields the following thermodynamic identity: 
\begin{equation}\label{tdid}
    n\nabla_a\left(\f{s}{n}\right) = \f{\nabla_a\rho}{T} - \left(\f{p_{\text{iso}} + \rho}{nT}\right)\nabla_an~. 
\end{equation}
In the above we used the Euler's relation, $\rho + p_{\text{iso}} = Ts + \mu n $.
Therefore, (\ref{BD2}) turns out to be
\begin{equation}\label{sl2}
    u^a\nabla_as = u^a\left\{\f{\nabla_a\rho}{T} - \left(\f{p_{\text{iso}} + \rho}{nT}\right)\nabla_an\right\} + \left(\f{s}{n}\right)u^a\nabla_an~.
\end{equation}
Also from the conservation $\nabla_aT^{ab} = 0$, if we expand $u_b\nabla_aT^{ab} =0$ using (\ref{emfull}), we obtain the relation
\begin{equation}\label{sl3}
    \f{u^a\nabla_a\rho}{T} = - \left[\left(\f{p_{\text{iso}}+\rho}{T}\right)\nabla_au^a +  \f{\pi^ab}{T}\nabla_au_b + \f{1}{T}q^au^b\nabla_bu_a + \f{\nabla_aq^a}{T}\right]~.
\end{equation}
We use (\ref{sl3}) in (\ref{sl2}) and substitute in (\ref{sl1}). Now, recall $\nabla_aN^a = 0$ where $N^a$ is given in (\ref{Na}). Thus we must have $\nabla_an^a = - \nabla_a\nu^a$. Using these we obtain
\begin{multline}\label{sdiv}
    \nabla_as^a =  q^a\left(\nabla_a\beta - \f{1}{T}u^b\nabla_bu_a\right) + \left(\beta - \f{1}{T}\right)\nabla_aq^a \\ 
    - \left[\lambda + \left(\f{s}{n} - \f{p_{\text{iso}}+\rho}{nT}\right)\right]\nabla_a\nu^a - \nu^a\nabla_a\lambda  - \f{\pi^{ab}}{T}\nabla_au_b~. 
\end{multline}

We want that to be positive definite to ensure the second law. This fixes the unknown parameters. Consider the choice, 
\begin{align}
    \beta &= \f{1}{T}~; \\ 
    \lambda &= \f{p+\rho}{nT} - \f{s}{n} = \f{\mu}{T}~.
\end{align}
Thus, two of the terms in (\ref{sdiv}) vanish. For the heat flux, consider 
\begin{equation}
    q^a = -\kappa h^{ab}\left[\nabla_bT + \dot{u}_bT\right]
\end{equation}
with the heat conductivity $\kappa \geq 0$ to ensure that the first term in (\ref{sdiv}) is positive. This is shown below. Using the choice of $\beta = 1/T$, one can show that  
\begin{equation}
    -q^a\f{1}{T^2}\left(\nabla_aT + T \dot{u}_a\right) = \f{1}{T^2}q^a\left\{-h_{ab}\left(\nabla^bT + T \dot{u}^b\right)\right\} = \f{q^aq_a}{\kappa T^2} \geq 0~.
\end{equation}
Similarly, consider the diffusion flux $\nu^a$ to have the form, 
\begin{equation}\label{andersonnu}
    \nu^a = - \sigma T^2h^{ab}\nabla_b\lambda~,
\end{equation}
where $\sigma \geq 0$ is the diffusion coefficient. Then, we can rewrite the term, 
\begin{equation}
    -\nu^a\nabla_a\lambda = \f{1}{\sigma T^2}\nu^a\left\{-\sigma T^2 h_{ab}\nabla^b\lambda\right\} = \f{\nu^a\nu_a}{\sigma T^2} \geq 0~.
\end{equation}
Finally, recalling (\ref{vdecomp}) one can easily check that
\begin{equation}
    -\f{1}{T}\pi^{ab}\sigma_{ab} = \f{1}{2\eta T}\pi^{ab}\left(-2\eta \sigma_{ab}\right) = \f{\pi^{ab}\pi_{ab}}{2\eta T}~.
\end{equation}
Putting all  together we obtain
\begin{equation}
    \nabla_as^a = \f{q^aq_a}{\kappa T^2} + \f{\nu^a\nu_a}{\sigma T^2} +  \f{\pi^{ab}\pi{ab}}{2\eta T}~.
\end{equation}
Hence, the second law of thermodynamics is ensured by construction if $ \f{\pi^{ab}\pi{ab}}{2\eta T}\geq 0$. Usually if $T$ is positive, then this is satisfied. However we found here that $T$ is negative. But since $\pi^{ab}\pi_{ab} =0$ in the present context, our construction is according to the second law of thermodynamics.
All the relations among the fluid and thermodynamic parameters are summarised in Eqs. (\ref{qeck}) -- (\ref{nu}).

\section{Evaluation of $N_0$ and $M_0$}\label{App01}
First, let us calculate  $N_0 = \nabla_e\Phi\nabla^e\Phi$, 
\begin{align}\label{N0}
    N_0 &= g^{ab}\nabla_a\Phi\nabla_b\Phi = g^{tt}\left(\nabla_t\Phi\right)^2 + g^{rr}\left(\nabla_r\Phi\right)^2 = \f{(z-f')^2 - 16p^2}{f}
\end{align}

Similarly, we can define $M_0 = \nabla^a\Phi\nabla^b\Phi\nabla_a\nabla_b\Phi$
\begin{align}\label{M0}
    M_0&= \left(g^{rr}\nabla_r\Phi\right)^2\nabla_r\nabla_r\Phi + \left(g^{tt}\nabla_t\Phi\right)^2\nabla_t\nabla_t\Phi \nonumber \\ 
    &+ 2~g^{rr}g^{tt}\nabla_r\Phi\nabla_t\Phi\nabla_r\nabla_t\Phi  
\end{align}

We will also need the connections in the $r,t$ coordinates to evaluate the higher order derivatives. The non-vanishing connections are given by, 
\begin{eqnarray}
    \Gamma^{r}_{rr} = - \f{f'}{2f} ~; \Gamma^{r}_{tt} = \f{ff'}{2} ~; \Gamma^t_{rt} =  \f{f'}{2f} 
\end{eqnarray}

Using these connections, we calculate the following higher derivatives of the scalar field:
\begin{eqnarray}
    \nabla_r\nabla_r\Phi = \f{(f')^2 - 2ff'' - zf'}{2f^2}~;\nabla_t\nabla_t\Phi = - \f{f'}{2}(z-f')~; \nabla_r\nabla_t\Phi = \f{2p~f'}{f} 
\end{eqnarray}

Now, we can compute $M_0$ through (\ref{M0}) using the above connections and higher derivatives,
\begin{align}
    M_0 = \nabla^a\Phi\nabla^b\Phi\nabla_a\nabla_b\Phi &= \left(z - f^{\prime}\right)^2\left[\f{1}{2}\left(\f{f^{\prime}}{f}\right)^2 - \f{f^{\prime \prime}}{f} - \f{z f^{\prime}}{2f^2}\right] + 16p^2\f{f^{\prime}}{2f^2}\left(z - f^{\prime}\right)~. 
\end{align}

\section{Calculation leading to (\ref{2choices})}\label{App 02}

We demand that $h_{ab}\nabla^bT = 0$ with $T = (-c_w/\kappa)\sqrt{-\nabla_i\Phi\nabla^i\Phi}$. Firstly, note that 
\begin{align*}
    h_{rr} &= g_{rr} + u_r~u_r = \f{16p^2}{f\left[16p^2 - (z-f^{\prime})^2\right]}
\end{align*}

Therefore, imposing the above condition gives us: 
\begin{align*}
   0 &= \kappa ~ \f{16p^2}{\left[16p^2 - (z-f^{\prime})^2\right]}\nabla_r\left[\f{-c_w}{\kappa}\sqrt{\f{16p^2 - (z-f^{\prime})^2}{f}}\right] \\ 
    &= \f{16p^2}{\left[16p^2 - (z-f^{\prime})^2\right]} \sqrt{\f{16p^2 - (z-f^{\prime})^2}{f}} \left\{-\nabla_r(\ln{\kappa}) + \nabla_r\left(\ln{\left[\sqrt{\f{16p^2 - (z-f^{\prime})^2}{f}}\right]}\right)\right\}
\end{align*}

Hence, we have
\begin{equation*}
    \f{16p^2}{\sqrt{f\left[16p^2 - (z-f^{\prime})^2\right]}}\left\{-\nabla_r(\ln{\kappa}) + \nabla_r\left(\ln{\left[\sqrt{\f{16p^2 - (z-f^{\prime})^2}{f}}\right]}\right)\right\} = 0
\end{equation*}

This gives us the two possibilities given by (\ref{2choices}).

\newpage
\bibliographystyle{IEEEtran}
\bibliography{bibl.bib}

\end{document}